# Accurate Gap Determination in Monolayer and Bilayer Graphene/*h*-BN Moiré Superlattices

Hakseong Kim,[1*] Nicolas Leconte,[2,3*] Bheema L. Chittari,[2] Kenji Watanabe,[4] Takashi Taniguchi,[4] Allan H. MacDonald,[3] Jeil Jung[2§] and Suyong Jung[1§]

[1]Korea Research Institute of Standards and Science, Daejeon 34113, Korea
[2]Department of Physics, University of Seoul, Seoul 02504, Korea
[3]Department of Physics, The University of Texas at Austin, Austin, TX 78712
[4]Advanced Materials Laboratory, National Institute for Materials Science, 1-1 Namiki, Tsukuba 305-0044, Japan

**High mobility single and few-layer graphene sheets are in many ways attractive as nanoelectronic circuit hosts but lack energy gaps, which are essential to the operation of field-effect transistors. One of the methods used to create gaps in the spectrum of graphene systems is to form long period moiré patterns by aligning the graphene and hexagonal boron nitride (*h*-BN) substrate lattices. Here, we use planar tunneling devices with thin *h*-BN barriers to obtain direct and accurate tunneling spectroscopy measurements of the energy gaps in single- and bi-layer graphene-*h*-BN superlattice structures at charge neutrality (first Dirac point) and at integer moiré band occupancies (second Dirac point, SDP) as a function of external electric and magnetic fields and the interface twist angle. In single-layer graphene we find, in agreement with previous work, that gaps are formed at neutrality and at the hole-doped SDP, but not at the electron-doped SDP. Both primary and secondary gaps can be determined accurately by extrapolating Landau fan patterns to**

§ To whom correspondence should be addressed: jeiljung@uos.ac.kr, syjung@kriss.re.kr
* These authors contributed equally to this work.

**zero magnetic field and are as large as ≈ 17 meV for devices in near perfect alignment. For bilayer graphene, we find that gaps occur only at charge neutrality where they can be modified by an external electric field. Tunneling signatures of in-gap states around neutrality suggest the development of edge modes related to topologically non-trivial valley projected bands due to the combination of an external electric field and moiré superlattice patterns.**

The sublattice symmetry of graphene forbids the energy gap needed for graphene-based electronic applications.[1] Breaking sublattice symmetry would directly generate gaps, but requires atomic scale structure control that is still out of reach. Several alternative approaches for engineering graphene energy gaps have therefore been proposed, including narrowing graphene channels to sub nanometer dimensions[2] and subjecting graphene to periodic potentials to generate gaps at the superlattice-zone boundary.[3–5] Here we explore gap engineering based on recent experimental advances in realizing two-dimensional vertical van der Waals heterostructures between graphene and *h*-BN lattices. It is known that minimizing the orientation difference between the two lattices maximizes the influence of *h*-BN on the graphene electronic structure.[6–11] If the aligned lattices were commensurate, gaps near 50 meV would be expected at charge neutrality.[12] At the same time, the difference in lattice constants implies that moiré patterns are formed when the lattices are aligned,[12–15] giving rise to Hofstadter butterfly effects in a magnetic field and inducing second-generation Dirac points (SDPs) at the corners of the graphene-*h*-BN superlattice Brillouin zones.[16,17] The question that follows is whether graphene superlattices generate energy-gaps at the SDPs, as expected for graphene under a periodic potential,[3,4,15,16] and, if so, how large these gaps can be.

In single-layer graphene, the emergence of the energy gaps at the FDP and the SDPs has been experimentally confirmed by multi-probe conductance,[8,9,16,18] magneto-optical spectroscopy[19] and angle-resolved photoemission spectroscopy (ARPES) .[20] However, the gap sizes inferred from different experiments have varied. It has been argued that a delicate interdependence between the atomic and electronic structures of graphene-*h*-BN heterostructures could affect energy-gap formation. For example, alteration of the gaps could take place upon encapsulating devices with another *h*-BN layer,[21] and application of external pressure in graphene on *h*-BN leads to a progressive increase of the gap at the FDP due to enhanced interlayer coupling.[18] Indeed recent ARPES measurements have established that sub-Angstrom changes of the interlayer distance between graphene and *h*-BN layers can induce a one-order swing.[20] Some of us, in an earlier theoretical work, had argued that relaxations of both graphene and *h*-BN lattices upon forming graphene-*h*-BN heterostructures and electron-electron interactions both play important roles in deciding the energy-gap size at charge neutrality.[11] Compared with the rather rich experimental and theoretical reports on the gap states in single-layer graphene, however, the experimental studies on the effect of moiré patterns in bilayer graphene are practically absent in the literature, and little is known about the influence of interactions with their underlying *h*-BN substrates.

Electron tunneling spectroscopy is a reliable and accurate metrology for analyzing the electronic structure of solid-state systems, and is especially useful for those with energy gaps, for example, superconductivity, charge- and spin-density waves, or quantum Hall effects.[22–24] By monitoring the tunneling current variation (differential conductance) with respect to energy (or tunneling bias voltage), the size of gaps and their positions with respect to the Fermi level ($E_F$) can be acquired straightforwardly and more directly than in temperature-dependent multi-probe transport or magneto-optical spectroscopy measurements. With a scanning tunneling spectroscopy measurement, Yankowitz *et al.* explicitly demonstrated the formation of SDPs in graphene-*h*-BN superstructures and tracked their evolution as functions of an external gate voltage and the period of graphene superlattice.[25] Despite the obvious advantages of tunneling spectroscopy for addressing gapped states, however, extracting accurate energy-gap sizes at the SDPs and even at the FDP has not been a focus, largely due to the ambiguity of locating the energy-gap boundaries in tunneling spectra,[26] a problem we will discuss later in detail.

In this report, we present tunneling spectroscopy measurements on single- and bi-layer graphene-*h*-BN superlattices with thin *h*-BN as a tunnel barrier fabricated using mechanical exfoliation and transfer approaches. The enhanced tunnel-junction stability of our graphene-*h*-BN heterostructures[27,28] allows us to identify the positions of SDPs, and to establish whether an energy gap is formed at both the FDP and the SDPs as functions of the twist angle; the size of graphene superlattices, external electric and magnetic fields, and the number of carbon layers. As a major advantage over previous reports addressing gaps,[8,9,17–20,29] we accurately track band edges by monitoring the evolution of Landau levels (LLs) that originate from energy-gap boundaries when gaps exist, or at band-crossing points in the absence of an energy gap. In single-layer graphene superlattices, the energy gap at the FDP becomes as large as $\approx 17$ meV at near-perfect alignment. The broken-inversion symmetry of graphene-*h*-BN superlattices causes electron-hole asymmetries in the electronic structures of single-layer graphene superlattices, opening an energy gap as large as $\approx 16$ meV at the SDP in the hole-doped region. No gap opens at the SDP in the electron-doped area. We observe that the gap at the SDP closes quickly as the twist-angle increases, while the gap size at the FDP decreases more slowly, and interpret the difference as being due to different microscopic origins of energy-gap formations in the two cases, *i.e.* that inversion symmetry breaking is sufficient at the FDP whereas the details of pseudospin mixing are crucial at SDPs.[30] In bilayer graphene superlattices, energy gaps fail to emerge at either the electron- and hole-doped SDPs because of a higher overall density of states and the presence of additional bands at mini-zone boundaries near the relevant energy. We establish that remote hopping contributions to the tight-binding Hamiltonian for intrinsic bilayer graphene play an essential role near the SDPs, and must be incorporated in the moiré superlattice Hamiltonian.[31] Near the FDP, we also observe tunneling signals that could be related to the edge modes of valley resolved topological states resulting from the simultaneous presence of an electric field and moiré patterns in bilayer superstructures,[32] whereas the electric-field induced energy gap dominates tunneling spectra at the charge neutrality point. All the key experimental observations are qualitatively understood within a simple single-particle picture of single- and bi-layer graphene-*h*-BN superlattices. A better quantitative agreement of the gap sizes with experiment should be achievable through the inclusion of many-body Coulomb interactions.

## Results

**Probing monolayer and bilayer graphene/*h*-BN Moiré superlattices.** The optical viewgraph and the schematics of the tunneling and multi-probe devices are shown in Fig. 1d. Firstly, we carefully select mechanically exfoliated graphene flakes with well-defined crystalline edges. For some devices, flakes with both single- and bi-layer regions are chosen in order to systematically investigate layer-dependence effects on graphene-*h*-BN superstructures with the same twist angle. Next, a thin *h*-BN flake with three to five layers and a thick graphite probe are sequentially transferred on top of the crystallographically aligned graphene-*h*-BN heterostructures to form planar graphene-*h*-BN tunnel junctions. To provide complementary information, we also fabricated edge-contacted multi-probe graphene devices, sharing both bottom $h$-BN/SiO$_2$ back-gate insulators and the graphene layer with tunneling devices (Fig. 1d). With edge-contacted devices, the presence of the graphene-$h$-BN superlattices are confirmed through multi-probe conductance measurements with additional conductance (or resistance) dips (peaks) (Fig. S1). We obtain the back-gate capacitance ($C_g$) defined by bottom $h$-BN and SiO$_2$ insulators from the Wannier-type fan diagram in high magnetic field measurements. With attained $C_g$ and $V_g$-spacings between the SDPs and the FDP, basic parameters of graphene-$h$-BN superlattices can be extracted such as the location of the SDP in energy with respect to the FDP, the size of graphene-$h$-BN superlattice, and the twist angle defined between graphene and bottom $h$-BN flakes (See Supporting Information). The twist angle between graphene and $h$-BN layers can be estimated from Raman measurements as well.[33] As displayed in Fig. S3, the Raman intensity and the width of the 2D peak are heavily influenced by the formation of a graphene-$h$-BN superlattice.[33] In our devices, the twist angles estimated from the Raman measurements are consistent with the values from multi-probe measurements, implying that our graphene superlattices have little disarray from nearby graphene ripples or bubbles.[34]

Figures 1e and 1g display the series of tunneling spectra, differential conductance ($G = dI/dV_b$) curves at $B = 0$ T from the single-layer (SL, Fig. 1e, sample **A**) and bi-layer (BL, Fig. 1f, sample **B**) graphene-$h$-BN superlattices. Both SL- and BL-graphene devices are fabricated with a flake where SL and BL regions coexist, and the misalignment angle with the underlying $h$-BN insulator $\theta$ is small ($\theta \leq 0.1°$). With these small twist angles, the graphene-$h$-BN superlattice defined in the single-layer region has SDPs located $\approx \pm 0.17$ eV away from the FDP and a moiré lattice constant estimated to be $\lambda_M \approx 13.9$ nm. At zero $V_g$, a $dI/dV_b$-dip feature relating to diminished density of states at the FDP is positioned at $V_b \approx 0$ mV for un-doped SL- and BL-graphene devices, as marked with orange triangles in Figs. 1e and 1f. Similar to our previous report,[28] the d$I$/d$V_b$-dip feature at the FDP shifts away from $V_b = 0$ mV as $V_g$ increases from $V_g = 0$ V to $V_g = \pm 10$ V in magnitude. In addition, the FDP d$I$/d$V_b$ feature in the bilayer (Sample **B**, Fig.1f) evolves into a 'U'-shaped gap that widens as the electric field-induced gap increases at non-zero $V_g$.[28] Here, the most noticeable differences, compared with strongly misaligned graphene-$h$-BN heterostructures[28] are the presence of additional d$I$/d$V_b$-dip structures, marked with red and blue triangles in Figs. 1e and 1f. The pair of secondary d$I$/d$V_b$ dips are equidistant at both sides of the FDP, and change their positions together with FDP for different $V_g$. The $V_g$-dependent characteristics confirm that these secondary d$I$/d$V_b$ features are linked to the electronic structures of the graphene-$h$-BN superlattices, particularly the SDPs at the mini-zone boundary.[27,35]

In the sample **A** (Fig. 1e), the d$I$/d$V_b$ dip positioned lower than the FDP in $V_b$, therefore formed in the hole-doped region, is conspicuous when compared to the d$I$/d$V_b$ dip located in the electron-doped area. Quite on the contrary, both d$I$/d$V_b$ dips in the sample **B** (Fig. 1f) for BL-$h$-BN superlattices have similar d$I$/d$V_b$ magnitude for electrons and holes. We note that the electron-hole asymmetry in graphene-$h$-BN superstructures can be traced to the destructive and constructive combinations for electrons and holes of the moiré pattern matrix elements due to local mass and virtual strain patterns of comparable magnitudes.[36] These obvious differences in the degree of electron-hole asymmetry between SL- and BL-$h$-BN superlattices are due to negligibly small virtual strains that couple the two low-energy carbon sites in bilayer graphene on $h$-BN.

Figures 2a and 2b show the two-dimensional representation of the tunneling spectra in the window of $V_g$ and $V_b$, which consist of 121-independent d$I$/d$V_b$ curves (Figs. 1e and 1f) from $V_g$ = -30 V to $V_g$ = 30 V with a spacing of $\Delta V_g$ = 0.5 V, respectively for the SL (sample **A**) and

BL (sample **B**)-*h*-BN superlattices.  In the sample **A** (Fig. 2a), the suppressed d$I$/d$V_b$ at the FDP

(marked with a dotted white arrow) and the secondary d$I$/d$V_b$ dip (marked with a dotted yellow

arrow) for the SDP in the hole-doped region are well recognized.  Along with those dark d$I$/d$V_b$

bands running diagonally across the gate map, an additional d$I$/d$V_b$ dip (marked with a dotted

green arrow) at the SDP in the electron-doped area is shown distinctly with a significantly

weakened visibility in the gate mapping (Fig.2a).  In contrast, as shown in Fig. 2b, the electron-

hole asymmetries become weaker in the BL-*h*-BN superlattice, featuring as a pair of d$I$/d$V_b$ dips

equidistantly spaced in $V_b$ around the electrostatically induced bilayer-graphene energy gap at the

FDP.  As discussed in our previous report,[28] the bilayer energy gap at the FDP is proportional to

the width of the suppressed d$I$/d$V_b$ band, which develops in the presence of an electric field

between carbon layers.[28]

Here, note that the suppressed visibility of the d$I$/d$V_b$ dips at the SDPs in the bilayer is close

to the d$I$/d$V_b$ at the SDP in the electron-doped region in the single layer.  In contrast, the d$I$/d$V_b$

dips at the FDP and the SDP in the hole-doped region of the SL-*h*-BN superlattice (Fig. 2a) have

d$I$/d$V_b$ signatures, just as clearly visible as the gap at the FDP in the BL-*h*-BN superlattice (Fig.

2b).  We find that the observed visibility differences in d$I$/d$V_b$ are related to the presence of energy gaps either at the FDP or at the SDPs in SL- and BL-*h*-BN superlattices.  Additionally,

the tunneling features in the planar graphene-*h*-BN tunnel devices predominantly reflect the

electronic structures at $E_F$, which is tunable through the capacitive couplings across the thin *h*-

BN tunnel insulator and the relatively thick bottom-gate *h*-BN layers.  Thus, the absence of observable variations in both the linear scaling of the FDP and the SDPs and the equidistance of the SDPs from the FDP confirm that the planar capacitive model is sufficient to explain the tunneling spectra, guaranteeing little variations in the electronic structure of graphene superlattices in our measurement $V_g$–$V_b$ window.

**Energy gaps in monolayer graphene/*h*-BN Moiré superlattices.** We now discuss in detail the electronic-structure variations and plausible energy-gap formations at the FDP and the SDPs at the mini-zone boundaries in graphene-*h*-BN superlattices. Although the strongly suppressed d$I$/d$V_b$ spectra suggest the presence of energy gaps at the FDP and the SDP in the hole-doped region in the SL-*h*-BN superlattice, the dataset at $B = 0$ T is not sufficient either as indisputable evidence for the presence of a gap or for quantitative energy-gap analyses. Indeed, a suppressed d$I$/d$V_b$ band can be observed at the FDP in completely misaligned graphene-*h*-BN heterostructures where no gap exists.[28] A low-valued d$I$/d$V_b$ band is also present at $V_b = 0$ mV (Figs. 2a and 2b), which is ascribed to the diminished phonon density of states and reduced electron tunneling rates in graphene and other low-dimensional systems.[28,37,38] Moreover, tunneling signals are often delineated by d$I$/d$V_b$ features like the strong d$I$/d$V_b$ peaks marked with blue arrows in Figs. 1e and 2a. Those resonance peaks are due to the relative crystalline-angle alignments of graphene and the top graphite probe, and are not directly related to the electronic structures of graphene-*h*-BN superlattices. Landau level spectroscopy, in turn, is proven to be the most efficient and accurate metrology for quantitative analyses of gapped states in graphene.[26] In the presence of a perpendicular magnetic field, charged carriers in graphene collapse into well-separated LLs and the lowest LLs for electrons and holes coincide with the energy-gap boundaries. Therefore, the accurate assessment of energy gaps can be directly inferred from monitoring LL evolutions as the external magnetic field varies.

Figure 2c displays the gate mapping of the sample **A**, SL-*h*-BN superlattice at $B = 1$ T, revealing the most direct and indisputable evidences for the energy-gap formations at the FDP and the SDP in the hole-doped region, and the absence of an energy gap at the SDP in the electron-doped area. First, we can easily identify suppressed d$I$/d$V_b$ bands where the FDP and the SDP in the hole-doped region exist at $B = 0$ T, as indicated with dotted white and yellow lines in Fig. 2c. Around these suppressed d$I$/d$V_b$ bands, a pair of LLs, each denoted as $LL_{0+}$ ($LL_{0+,SDP}$) and $LL_{0-}$ ($LL_{0-, SDP}$) exist with other LLs with higher LL indices, $N$. In comparison, a d$I$/d$V_b$ peak forms exactly at the SDP in the electron-doped region with a pair of dark lines with low d$I$/d$V_b$ in value. The d$I$/d$V_b$ peak, representing the lowest LL at the SDP in the electron-doped area is marked with a dotted green arrow and one of the gaps in the spectra is guided with a dotted white line. The developments of LLs at both Dirac points can be traced to magnetic fields as low as $B = 0.1$ T.

Figure 2d shows the gate mapping of the sample **B**, BL-*h*-BN superlattice, measured at $B = 1$ T. In contrast with the SL-*h*-BN superlattice, aforementioned spectra features relating to energy-gap formations such as suppressed d$I$/d$V_b$ band paired with a couple of LLs are completely absent. The lowest-indexed LL, which is supposedly formed at Dirac points without a gap is not developed in BL-*h*-BN superlattices either. Instead, as marked with dotted yellow and green arrows in Fig. 2d, d$I$/d$V_b$ dips are eminent at both the SDPs with far lower visibility than the ones for the gaps in the SL-*h*-BN superlattice. Around the d$I$/d$V_b$ dips, a secondary pair of LLs is observed at the lower $V_b$ for the SDP in the hole-doping region and at the higher $V_b$ for the SDP in the electron-doped area. All these observations point out that the energy gaps at both SDPs are not completely developed due to additional accessible bands around the BL-*h*-BN superlattice mini-zone boundaries, which will be further confirmed with detailed experimental results and theoretical calculations. We also defer the experimental and theoretical discussions on the electronic structures at the charge-neutrality point to a later part in this manuscript.

**Landau-level spectroscopy.** We are able to investigate the energy-gap formations and the states around both Dirac points in more detail by utilizing LL spectroscopy through fine control of external magnetic fields at a fixed $V_g$. Figure 3a shows the LL-fan diagram from the SL-$h$-BN superlattice (sample **A**) with a spacing of $B = 0.1$ T. The gate voltage is adjusted at $V_g = -20$ V to better resolve the electronic structures at the FDP and both SDPs. The presence of energy gaps, represented by the bands of suppressed d$I$/d$V_b$, is easily identified with the LLs that arise from the boundaries of the gaps at the FDP (delineated with a dotted red box) and the SDP in the hole-doped region (marked with a dotted green box). Note that, at high magnetic fields, the tunneling associated to LLs and energy gaps often interfere with other features with different slopes as marked with dotted white lines in Fig. 3a. Those resonances are due to the relative positions of LLs and electronic band structures of a graphite probe and graphene-$h$-BN superlattices in energy-momentum space.[39] Additionally, the pre-split LLs at $B \approx 0$ T become fully split into four well-separated LLs with newly developed gaps due to broken spin degeneracy and electron-electron interactions, allowing to quantitatively measure the gaps resulting from the broken inversion symmetry of graphene-$h$-BN superlattices. Figures 3b–3d show the high-resolution d$I$/d$V_b$ spectra from the sample **A** at the FDP (Fig. 3b), the SDP in the hole-doped region (Fig. 3c), and the SDP in the electron-doped area (Fig. 3d), respectively. The suppressed d$I$/d$V_b$ bands with a pair of LLs around evidence the energy gap at the FDP (Fig. 3b) and the SDP in the hole-doped region (Fig. 3c) in SL-$h$-BN superlattices. Conversely, the LL ($LL_{0,\,SDP}$) developed right at the SDP (Fig. 3d) proves that no energy gap is formed at the mini-zone boundary in the electron-doped area.

In our planar graphene-$h$-BN tunnel devices, $E_F$ of graphene does not change its position in energy even with varying $V_b$, when electro-statically induced charged carriers are occupying the LLs. Accordingly, the high conductance d$I$/d$V_b$ bands representing each LL extend their width in $V_b$ as external $B$ increases. Thick yellow and orange ticks in Figs. 3a–3d indicate how much $V_b$ is required to fill the LLs with the filling factor (FF) of either two or four depending on whether valley symmetry is broken or not. We consider that spin degeneracy is not lifted at lower $B$. Thus, accurate energy-gap assessment is obtained from extrapolating each LL-peak position to $B = 0$ T and extracting $V_b$ spacing for an energy gap. The white circles in Figs. 3b and 3c indicate LL-peak positions at different $B$ at the FDP and the SDP, and the energy gaps are estimated to be $(16.2 \pm 0.1)$ mV at the FDP and $(15.6 \pm 1.0)$ mV at the SDP in the hole-doped region for the SL-$h$-BN superlattice in near perfect alignments ($\theta \leq 0.1°$). Here, we point out that the $V_b$-spacings measured in our graphene-$h$-BN tunnel junctions can be directly related to the actual energy bandgaps because of the large geometric capacitance of planar tunnel junctions and the absence of in-gap states.[28,40]

**Energy gaps *vs*. twist angles.** The energy gaps induced from the broken inversion symmetries of SL-$h$-BN superlattices are expected to be sensitive to the twist angles of graphene and underlying $h$-BN substrates,[30] and our measurements confirm the expectations. In total, we have measured eight SL-$h$-BN superlattice devices with different twist angles, and the extracted energy gaps are summarized in Fig. 4. There are a few intriguing points to note in the energy-gap evolutions at different twist angles. Firstly, energy gaps at the SDP are as big as those at the FDP in size at $\approx 16$ meV for closely aligned SL-$h$-BN superlattices, but close out much faster than the FDP gap as the twist angle increases. For SL-$h$-BN superlattices with a twist angle larger than $0.6°$, SDP energy gaps close out almost completely and d$I$/d$V_b$ features of a gap are practically absent. In contrast, the energy gap at the FDP shrinks with a rather moderate rate for increasing twist angle. Even for the SL-$h$-BN superlattice with a twist angle of $1.1°$, the gap is estimated to be as large as $(11.7 \pm 0.2)$ meV.

The obvious differences at the FDP and SDP highlight the microscopic-scale differences of the energy-gap formation mechanisms at the FDP and the SDP in SL-$h$-BN superlattices. The closed blue squares and red circles in the inset of Fig. 4 respectively represent theoretically expected energy gaps at the FDP and the SDP in hole-doped regions, with a consideration of fully relaxed graphene and $h$-BN lattices without electron-electron interactions. Here, we point out that simple single-particle considerations are sufficient enough to qualitatively explain most of the experimentally observed characteristics at both Dirac points. Expected gaps at both the FDP and the SDP are close to $\approx 8$ meV in size at a zero twist angle, and the fast closure of SDP gaps at $0.6°$ and larger twist angles is confirmed, as well as the modest decrease of the energy-gap sizes for the FDP. Note that the experimentally observed energy gaps are consistently larger than theoretically obtained ones by approximately a factor of two. We attribute the approximately two-fold increase of the experimental data to substantially enhanced electron-electron interaction effects, which are expected in high quality graphene-$h$-BN heterostructures.[41] Within a single-particle picture, we can further calculate the LL evolutions near the gaps at both Dirac points. Numerically generated LL-fan diagrams are presented for the FDP in Fig. 3f, the SDPs in the hole- and electron-doped regions in Figs. 3g and 3h, and the agreement with the experimental data is more than satisfactory. The valley degeneracy at both SDPs is considered $g_v = 2$, same as the degeneracy at the FDP, as experimentally confirmed with the orange (FF = 4) and the yellow (FF = 2) tick marks in Figs. 3a–3d.[42]

**Electronic structures of bilayer graphene/*h*-BN Moiré superlattices.** Next, we switch our discussions to the electronic structure variations at the charge neutrality point and the mini-zone boundaries in BL-*h*-BN superlattices. As shown in Figs. 2b and 2d, the most radical differences of BL-*h*-BN superlattice structures when compared with SL-*h*-BN devices are the lack of the fully-formed energy gaps with moderate electron-hole asymmetries at both SDPs. Figure 5a shows the LL-fan diagram from the BL-*h*-BN superlattice (sample **B**) with a spacing of $B = 0.1$ T at a fixed $V_g = 10$ V. It is clear that no fully-formed energy gaps develop at the SDPs, with the electrostatically induced gap at the charge neutrality point at $V_b \approx$ -120 mV and $V_g = 10$ V. The absence of the energy gaps at the SDPs is further confirmed in the high-resolution d$I$/d$V_b$ mappings taken at the SDP in the hole-doped region (Fig. 5b) at $V_g = 10$ V, and the SDP in the electron-doped area (Fig. 5c) at $V_g = $ -20 V, respectively. Around the weakly suppressed d$I$/d$V_b$ bands at the SDPs, secondary sets of LLs are present at lower $V_b$ for the SDP in the hole-doping region and higher $V_b$ for the SDP in the electron-doped area. At $B = 1$ T, the secondary LLs start interfering with the gapped structures at the SDP and completely overtake them at higher B fields.

We numerically calculate the electronic band structure of BL-*h*-BN superlattices (Fig. 1c), and the LL-fan diagrams around the FDP and the SDPs (Figs. 5e–5h). We find that the simplest model accounting only for interaction effects between the *h*-BN and the bottom graphene layer is sufficient to capture most of the experimental observations in the LL-fan diagrams. A moderately advanced model considering the interactions between the *h*-BN and the top graphene layer is found to slightly facilitate the electron-hole symmetries by enhancing tunnel-feature visibility at the SDP in the electron-doped region. A full DFT model accounting for the coupling with the substrate would likely improve the agreement even more, but a detailed analysis is beyond the scope of the current study. As for the tunneling features at the FDP, however, the electrostatic interactions with the moiré patterns in BL-$h$-BN superlattices seems to be critical to understand experimental observations, as discussed below.

**Mid-gap modes in bilayer graphene/$h$-BN moiré superlattices.** Probing the electronic structures at the charge neutrality point (FDP) in BL-$h$-BN superlattices is not as straightforward as the previously discussed tunneling features at the SDPs and those for single-layer graphene superstructures. As shown in the gate mappings (Figs. 2b and 2d) and the LL-fan diagram in Fig. 5a, the electric-field induced energy gap dominates the tunneling features since there exists a single set of electron- and hole-energy bands around the charge neutrality point in bilayer systems. Quite interestingly, however, we observe a faint-but-clear pair of d$I$/d$V_b$ peaks developing inside the electric-field induced gap in both positive and negative $V_b$ regions, as marked with red arrows in Figs. 2b and 2d. These d$I$/d$V_b$ peaks evolve with a different angle across the gate mappings when compared with the slopes of electric-field induced energy-gap boundaries, suggesting that the observed d$I$/d$V_b$ peaks inside the gap are not directly related to the layer-symmetry breaking in bilayer graphene. In addition, the mid-gap structures do not change their positions in $V_b$ while varying $B$, as marked with a dotted red line in the high-resolution d$I$/d$V_b$ mapping in Fig. 5d. In total, we have measured three BL-$h$-BN superlattices and all three devices reveal similar d$I$/d$V_b$ features inside electric-field induced energy gaps.

Figure 6 shows the high-resolution $dI/dV_b$ spectra map from the additional BL-*h*-BN superlattice device with a twist angle of 0.65°. The mid-gap feature developed inside the electrostatically induced gaps is marked with a dotted red arrow. It is interesting to point out that the spectra are much clearer in the sample **C** (Fig. 6) when compared with those in the sample **B** (Fig. 2b), indicating that the planar graphite-tunnel *h*-BN-BL-*h*-BN superlattice tunnel junction in the sample **C** is less susceptible to momentum-relaxation tunneling events from mechanical deformations such as bubbles and structural defects in the junction. The strong resonance peaks, appearing as negative $dI/dV_b$ bands (blue arrows in Fig. 6), further support that the relative crystalline-angle alignments of graphene and the top graphite probe are well preserved in the sample **C**. These intriguing in-gap tunneling signals could be consistent with the picture of the edge modes associated with the low energy Chern bands expected from the simultaneous presence of the moiré superlattices and vertical electric fields in BL-*h*-BN superlattices being pushed into the gap.[32] In Fig. S8, we show the Berry curvature plots of the low energy moiré bands of BL-*h*-BN superlattice that gives rise to either zero or finite Chern number $C = \pm 2$ depending on the sign of interlayer potential difference.

**Discussion**

We have carried out a careful analysis of the electronic structure modifications in graphene-*h*-BN superlattice structures with varying twist angle, external electric and magnetic fields, and the number of carbon layers. Benefited by the much improved tunnel-junction stability with thin *h*-BN as a tunnel barrier, the energy gaps formed both at the FDP and the SDP in the hole-doped region in SL-*h*-BN superlattices are accurately assessed by high-resolution LL spectroscopies. The absence of the gap around the electron-doped SDP is confirmed. In BL-*h*-BN superlattices, the electron-hole asymmetries are significantly moderated with an energy gap not fully developed at both SDPs. We also find out experimental signatures of the mid-gap modes at the FDP due to the simultaneous presence of an electric field and moiré patterns in BL-$h$-BN superlattices. All the key experimental observations are qualitatively understood within a single-particle picture.

## Methods

**Device Fabrication.** Our tunneling and multi-probe devices are fabricated with multiple steps of dry-transfer method. First, bottom thick $h$-BN (> 50 nm) is mechanically exfoliated from high-quality single crystals and transferred onto thermally grown 90 nm thick $SiO_2$ on Si. We carefully examine the $h$-BN flakes with well-defined crystallographic edges under bright- and dark-filtered optical microscope and an atomic force microscope to ensure surface cleanness. Then, either single- or bi-layer graphene flakes are mechanically exfoliated on the stacks of PMMA (poly(methyl Methacrylate)) and water soluble PSS (poly-styrene sulfonic) layers, and transferred on top of to the pre-located $h$-BN flakes with a micro-manipulating transfer module equipped with a rotation stage for accurate twist-angle alignments. We remove the PMMA film in warm acetone for an hour and annealed the samples at 350 $^o$C for 7 hours in a mixture of Ar : $H_2$ = 9 :1 to guarantee residue-free graphene surfaces and reduce graphene bubbles formed at the graphene-$h$-BN interface. Transferred graphene flakes are then patterned into graphene-ribbon structures whose widths are in the range of a couple of microns, utilizing conventional electron-beam lithography and $O_2$-based dry-etching procedures. Next, either thin (three to five layers) or thick (> 50 nm) $h$-BN flakes are respectively transferred on top of the tunneling or multi-probe graphene/$h$-BN heterostructures with the aforementioned dry-transfer methods with PMMA/PSS polymer stacks. During the top *h*-BN transfers, we intentionally misalign the twist angles of the top *h*-BN flakes and underlying graphene/*h*-BN superstructures. After each transfer, we meticulously repeat the procedures of removing PMMA films in the solvent and annealing the samples in an Ar-H$_2$ environment at elevated temperature. For planar tunneling devices, the graphite flake is transferred on top of the thin *h*-BN tunnel insulator, which covers the pre-patterned single- and bi-layer graphene/*h*-BN moiré superlattices, while final devices are completed through an electron-beam lithography and metal (Ti ≈ 5 nm / Au ≈ 60 nm) lift-off process. For edge-contacted multi-probe devices, separate electron-beam lithography and dry-etching procedures are performed to pattern graphene Hall-bar structures in a single-layer region, and electric contacts are made with metal evaporations of Cr ( ≈ 5 nm) and Au (≈ 60 nm) films.

**Figure Captions**

**Figure 1| Schematic and energy-band diagrams of graphene-*h*-BN superlattice structures.**
**a**, Schematic of single-layer graphene-*h*-BN superlattices at a twist angle $\theta$ with detailed atomic structure configurations of BA, AB, and AA with carbon, nitrogen and boron atoms. The hexagonal superlattice structure with a moiré wavelength $\lambda_M$ is marked with a green hexagon. **b,c** Energy band diagrams of (**b**) single- and (**c**) bi-layer graphene-*h*-BN superlattices at $\theta = 0°$. For bilayer graphene, no electric-field induced energy gap is considered for calculations. **d**, Optical viewgraph and schematics of graphene-*h*-BN heterostructures for a planar tunnel device and an edge-contacted multi-probe device. Both devices share a graphene and a bottom *h*-BN insulator. Red arrows indicate current flows for tunneling and quantum Hall devices. Scale bar is 10 μm. **e,f**, d$I$/d$V_b$ spectra in zero magnetic field as a function of back-gate voltage in steps of $\Delta V_g = 2$ V for the single (sample **A**, **e**) and bilayer (sample **B**, **f**) superstructures. Twist angle for both devices is $\theta < 0.1°$. Inverted orange triangles mark the FDP at charge-neutrality point, and red and blue ones indicate the SDPs for hole- and electron-doped regions, respectively.

**Figure 2| Two-dimensional display of d$I$/d$V_b$ spectra for single- and bi-layer-$h$-BN superlattices. a,b,** d$I$/d$V_b$ gate maps comprised with 121-independent spectrum measured at varying $V_g$ in steps of $\Delta V_g = 0.5$ V at $B = 0$ T for the sample **A** (**a**) and the sample **B** (**b**). The positions for the FDP is marked with a dotted white arrow, and those for the SDPs in hole- and electron-doped regions are indicated with dotted yellow and green arrows. **c,d,** d$I$/d$V_b$ gate maps at $B = 1$ T for the sample **A** (**c**) and the sample **B** (**d**). Each Landau level (LL) is labeled with a LL-index N ($LL_N$), and dotted arrows mark the same locations of the FDP and the SDPs as those in Figs. 2**a** and 2**b**.

**Figure 3| LL-fan diagrams from the SL-$h$-BN superlattice. a,** Two-dimensional display of d$I$/d$V_b$ tunneling spectra from the sample **A** at a fixed $V_g$ = -20 V and $T$ = 4 K with a varying external magnetic field up to $B$ = 5 T, revealing series of LLs originated from either the FDP or the SDP in the hole-doped region. (**b–d**) High resolution LL-fan diagrams focused on the FDP (**b**), the hole-doped SDP (**c**), and the electron-doped SDP (**d**) regions. Each graph is obtained from the zoomed-in areas, marked with dotted boxes in Fig. 3**a**. Orange and yellow tick marks indicate how much $V_b$ is required to fill the LLs with a filling factor FF = 4 and FF = 2, respectively. White circles mark LL-peak positions from Lorentzian fittings and energy gaps (marked with yellow arrows) are obtained from extrapolating LL-peak positions in $V_b$ to $B = 0$ T. (**e–h**) Numerically simulated LL-fan diagrams. Our $B$-field dependent tunneling spectroscopy shows the structure of the bands near the SDP. Qualitative agreement between the single-particle theory and experiments is found, while some differences in the slopes of the tunneling gaps can arise due to electrostatic charging effects of the LL.

**Figure 4| Energy gaps in SL-$h$-BN superlattices.** Energy gaps developed at the FDP (blue squares) and the SDP (red circles) in the hole-doped region in SL-$h$-BN superlattices at different twist angles. (inset) Numerically calculated energy gaps at the primary Dirac point in SL-$h$-BN superlattices with a consideration of fully relaxed graphene and $h$-BN lattices within a single-particle picture. We observe a fast closure of the secondary Dirac point energy gap that is independent of strain effects due to lattice relaxations.

**Figure 5| LL-fan diagrams from the BL-$h$-BN superlattice. a,** Two-dimensional display of d$I$/d$V_b$ tunneling spectra from the sample **B** at a fixed $V_g$ = 10 V and $T$ = 4 K with a varying external magnetic field up to $B$ = 5 T. The electro-statistically induced energy gap at the FDP is shown as a vertical dark band at $V_b \approx$ -125 mV, accompanying with series of LLs originated from the charge neutrality point in the bilayer graphene. (**b–d**) High resolution LL-fan diagrams focused on the hole-doped SDP (**b**) at $V_g$ = 10 V, the electron-doped SDP (**c**) at $V_g$ = -20 V, and the FDP (**d**) at $V_g$ = 30 V. Back-gate voltages are chosen for each region to better resolve the not-fully-formed energy gaps (marked with yellow arrows) and the LLs from nearby energy bands (marked with dotted white lines) at both the SDPs. The dotted red arrow (**d**) mark the additional d$I$/d$V_b$ peak inside the electric-field induced energy gap at the FDP. (**e–h**) Numerically

simulated LL-fan diagrams. The simulated diagrams capture the SDP band features that are in qualitative agreement with experiments. Some discrepancies in the slopes for the gaps can be from electrostatic charging effects due to filling of Landau levels which are not considered in our theoretical model.

**Figure 6| Electronic structures of BL-*h*-BN superlattices at the charge neutrality point.** High resolution d$I$/d$V_b$ gate map for the sample **C** with a twist angle 0.65°, focused on the d$I$/d$V_b$ tunneling features (marked with red arrows) inside the electric-field induced energy gap in BL-*h*-BN superlattices. Blue arrows mark the resonance peaks, appeared as negative d$I$/d$V_b$ bands in the gate map.

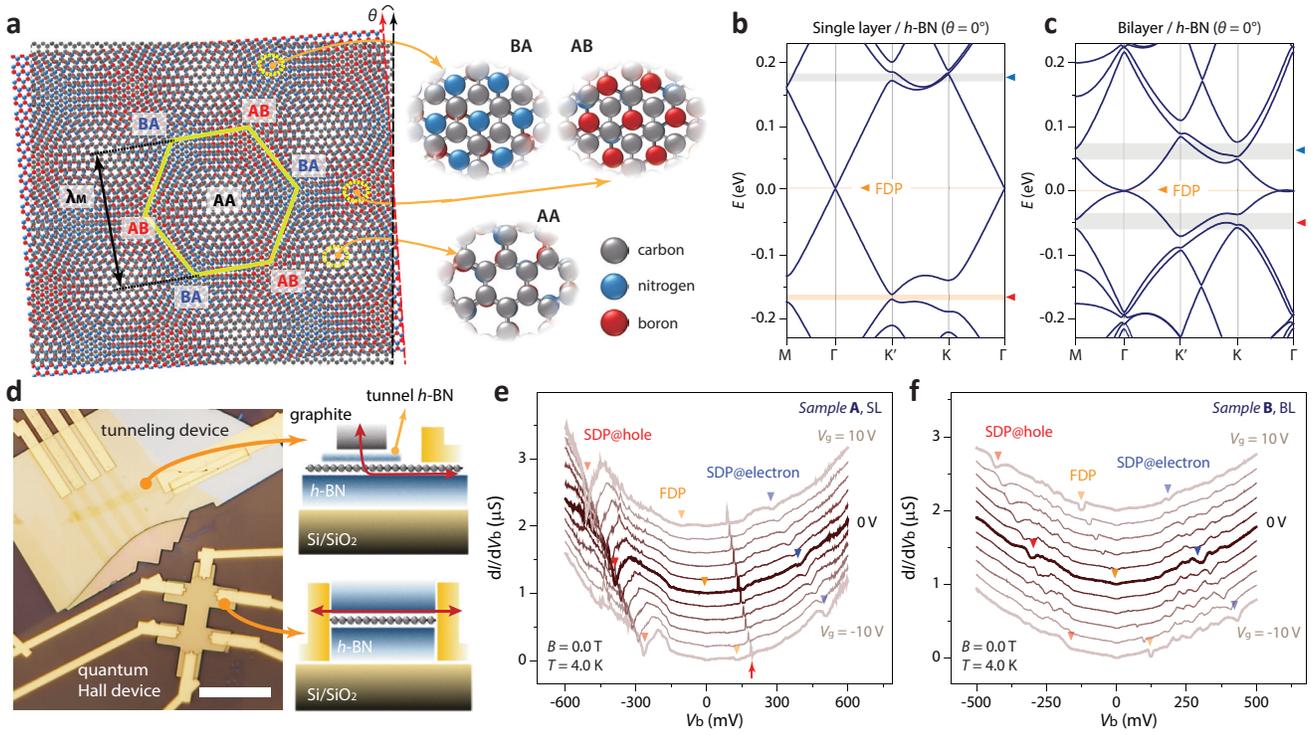

**Figure 1| Schematic and energy-band diagrams of graphene-*h*-BN superlattice structures. a**, Schematic of single-layer graphene-*h*-BN superlattices at a twist angle *θ* with detailed atomic structure configurations of BA, AB, and AA with carbon, nitrogen and boron atoms. The hexagonal superlattice structure with a moiré wavelength *λ*~M~ is marked with a green hexagon. **b,c**, Energy band diagrams of (**b**) single- and (**c**) bi-layer graphene-*h*-BN superlattices at *θ* = 0°. For bilayer graphene, no electric-field induced energy gap is considered for calculations. **d**, Optical viewgraph and schematics of graphene-*h*-BN heterostructures for a planar tunnel device and an edge-contacted multi-probe device. Both devices share a graphene and a bottom *h*-BN insulator. Red arrows indicate current flows for tunneling and quantum Hall devices. Scale bar is 10 μm. **e,f**, d*I*/d*V*~b~ spectra in zero magnetic field as a function of back-gate voltage in steps of Δ*V*~g~ = 2 V for the single (sample **A**, **e**) and bilayer (sample **B**, **f**) superstructures. Twist angle for both devices *θ* < 0.1°. Inverted orange triangles mark the FDP at charge-neutrality point, and red and blue ones indicate the SDPs for hole- and electron-doped regions, respectively.

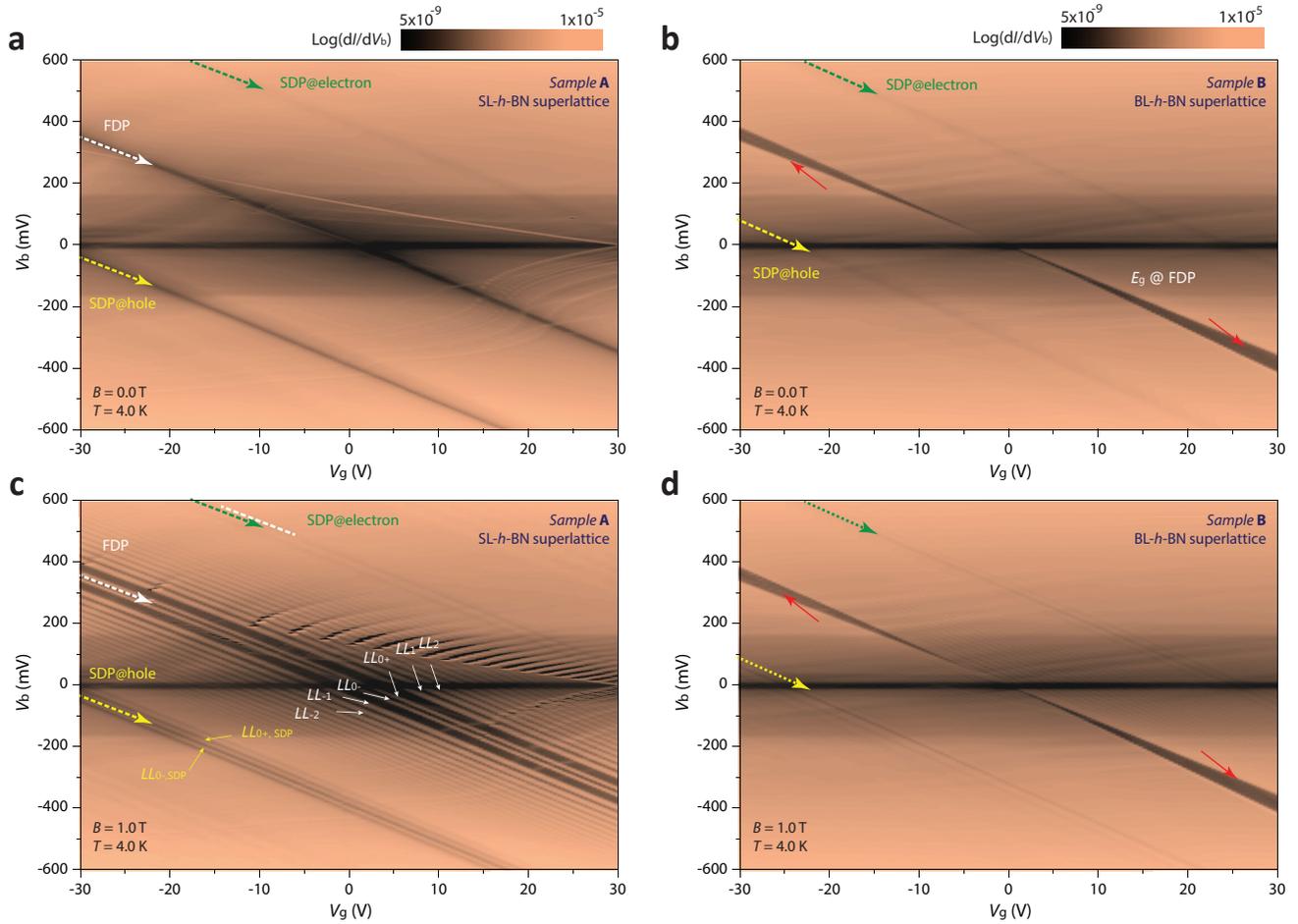

**Figure 2| Two-dimensional display of d$I$/d$V_b$ spectra for single- and bi-layer-*h*-BN superlattices.**
**a**,**b**, d$I$/d$V_b$ gate maps comprised of 121-independent spectrum measured at varying $V_g$ in steps of $\Delta V_g = 0.5$ V at $B = 0$ T for the sample **A** (**a**) and the sample **B** (**b**). The positions for the FDP is marked with a dotted white arrow, and those for the SDPs in hole- and electron-doped regions are indicated with dotted yellow and green arrows. **c**,**d**, d$I$/d$V_b$ gate maps at $B = 1$ T for the sample **A** (**c**) and the sample **B** (**d**). Each Landau level (LL) is labeled with a LL-index N ($LL_N$), and dotted arrows mark the same locations of the FDP and the SDPs as those in Figs. 2**a** and 2**b**.

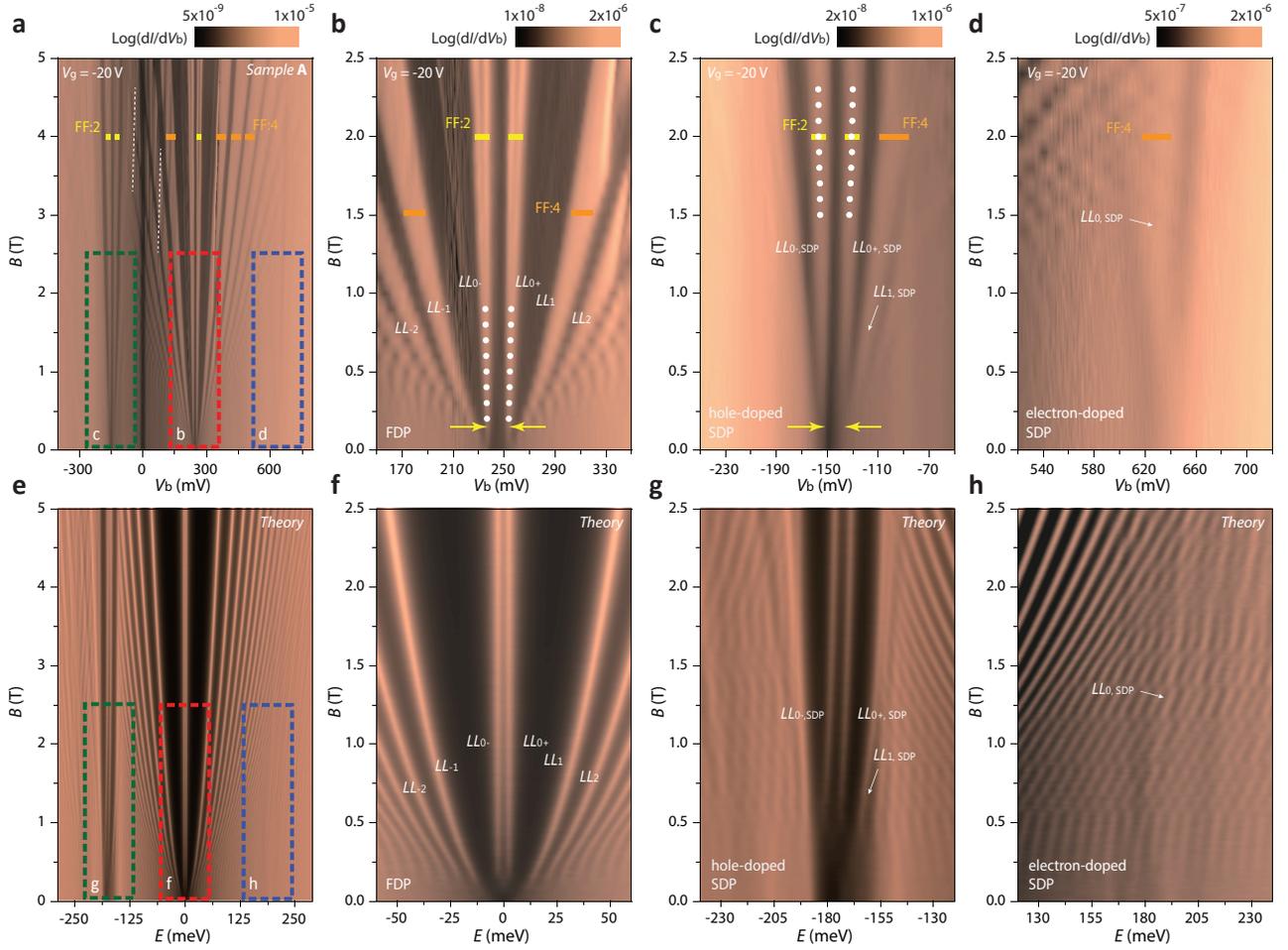

**Figure 3 | LL-fan diagrams from the SL-*h*-BN superlattice. a**, Two-dimensional display of d*I*/d*V*_b tunneling spectra from the sample **A** at a fixed $V_g$ = -20 V and $T$ = 4 K with a varying external magentic field up to $B$ = 5 T, revealing series of LLs originated from either the FDP or the SDP in the hole-doped region. (**b–d**) High resolution LL-fan diagrams focused on the FDP (**b**), the hole-doped SDP (**c**), and the electron-doped SDP (**d**) regions. Each graph is obtained from the zoomed-in areas, marked with dotted boxes in Fig. 3**a**. Orange and yellow tick marks indicate how much $V_b$ is required to fill the LLs with a filling factor FF = 4 and FF = 2, respectively. White circles mark LL-peak positions from Lorentzian fittings and energy gaps (marked with yellow arrows) are obtained from extrapolating LL-peak positions in $V_b$ to $B$ = 0 T. (**e–h**) Numerically simulated LL-fan diagrams. Our $B$-field dependent tunneling spectroscopy shows the structure of the bands near the SDP. Qualitative agreement between the single-particle theory and experiments is found, while some differences in the slopes of the tunneling gaps can arise due to electrostatic charging effects of the LL.

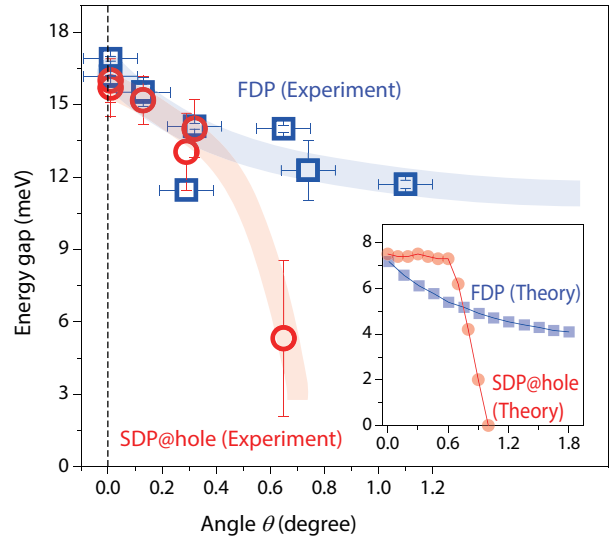

**Figure 4| Energy gaps in SL-*h*-BN superlattices.** Energy gaps developed at the FDP (blue squares) and the SDP (red circles) in the hole-doped region in SL-*h*-BN superlattices at different twist angles. (inset) Numerically calculated energy gaps at the primary Dirac point in SL-*h*-BN superlattices with a consideration of fully relaxed graphene and *h*-BN lattices within a single-particle picture. We observe a fast closure of the secondary Dirac point energy gap that is independentr of stain effects due to lattice relaxations.

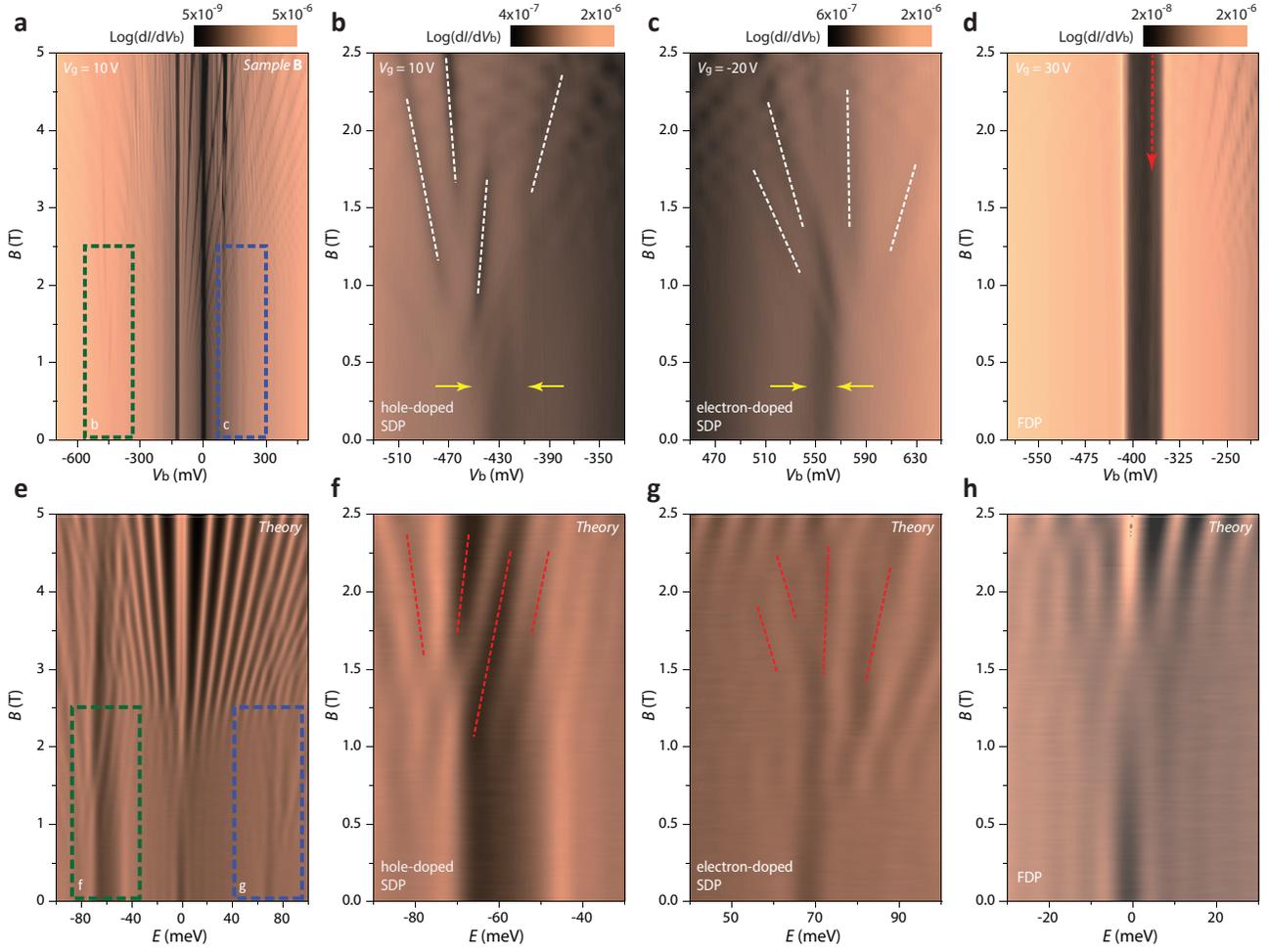

**Figure 5| LL-fan diagrams from the BL-*h*-BN superlattice. a**, Two-dimensional display of d*I*/d*V*ₑ tunneling spectra from the sample **B** at a fixed $V_g = 10$ V and $T = 4$ K with a varying external magentic field up to $B = 5$ T. The electro-statistically induced energy gap at the FDP is shown as a vertical dark band at $V_b \approx$ -125 mV, accompanying with series of LLs originated from the charge neutrality point in the bilayer graphene. (**b–d**) High resolution LL-fan diagrams focused on the hole-doped SDP (**b**) at $V_g = 10$ V, the electron-doped SDP (**c**) at $V_g = -20$ V, and the FDP (**d**) at $V_g = 30$. Back-gate voltages are chosen for each region to better resolve the not-fully-formed energy gaps (marked with yellow arrows) and the LLs from nearby energy bands (marked with dotted white lines) at both the SDPs. The dotted red arrow (**d**) mark the additional d*I*/d*V*ₑ peak inside the electric-field induced energy gap at the FDP. (**e–h**) Numerically simulated LL-fan diagrams. The simulated diagrams capture the SDP band features that are in qualitative agreement with experiments. Some discrepancies in the slopes for the gaps can be from electrostatic charging effects due to filling of Landau levels which are not considered in our theoretical model.

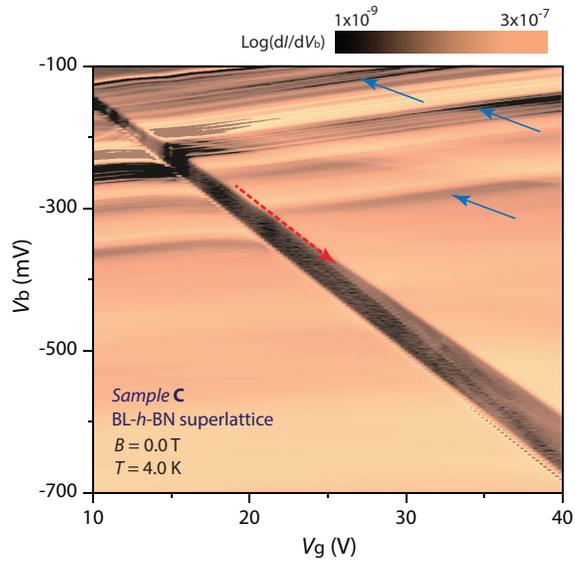

**Figure 6| Electronic structures of BL-*h*-BN superlattices at the charge neutrality point.** High resolution d$I$/d$V_b$ gate map for the sample **C** with a twist angle 0.65°, focused on the d$I$/d$V_b$ tunneling features (marked with red arrows) inside the electric-field induced energy gap in BL-*h*-BN superlattices. Blue arrows mark the resonance peaks, appeared as negative d$I$/d$V_b$ bands in the gate map.